\documentclass[10pt]{article}
\usepackage{amsfonts}
\usepackage{mathrsfs}
\usepackage{amsmath}	
\usepackage{cite}
\usepackage{graphicx}
\usepackage{rotating}
\usepackage{hyperref}
\usepackage{verbatim}
\usepackage[all]{xy}

\csname @addtoreset\endcsname{equation}{section}
\newcommand{\beq}{\begin{equation}}
\newcommand{\eeq}{\end{equation}}
\newcommand{\bea}{\begin{eqnarray}}
\newcommand{\eea}{\end{eqnarray}}
\newcommand{\ba}{\begin{array}}
\newcommand{\ea}{\end{array}}
\newcommand{\bit}{\begin{itemize}}
\newcommand{\eit}{\end{itemize}}
\newcommand{\nn}{\nonumber}

\newcommand{\mezzo}{\frac{1}{2}}
\newcommand{\complesso}{{\ \hbox{{\rm I}\kern-.6em\hbox{\bf C}}}}
\newcommand{\reale}{{\hbox{{\rm I}\kern-.2em\hbox{\rm R}}}}
\newcommand{\uno}{ \,  \raisebox{+0.14em}{{\hbox{{\rm \scriptsize ]}} \raisebox{-0.2em}{\kern-.8em\hbox{1}}}} \, }  

\newcommand{\p}{\partial}

\renewcommand{\a}{\alpha}
\renewcommand{\b}{\beta}
\newcommand{\g}{\gamma}

\newcommand{\G}{\Gamma}
\renewcommand{\d}{\delta}
\newcommand{\D}{\Delta}

\renewcommand{\k}{\kappa}
\renewcommand{\l}{\lambda}
\renewcommand{\L}{\Lambda}

\newcommand{\m}{\mu}

\newcommand{\n}{\nu}
\renewcommand{\r}{\rho}
\newcommand{\s}{\sigma}

\newcommand{\z}{\zeta}

\newcommand{\om}{\omega}
\newcommand{\Om}{\Omega}

\begin{document}


\begin{titlepage}
\begin{flushright}
CECS-PHY-15/06
\end{flushright}
\vspace{2.5cm}
\begin{center}
\renewcommand{\thefootnote}{\fnsymbol{footnote}}
{\huge \bf Magnetised Kerr/CFT correspondence}
\vskip 30mm
{\large {Marco Astorino\footnote{marco.astorino@gmail.com}}}\\
\renewcommand{\thefootnote}{\arabic{footnote}}
\setcounter{footnote}{0}
\vskip 10mm
{\small \textit{
Centro de Estudios Cient\'{\i}ficos (CECs), Valdivia,\\ 
Chile\\}
}
\end{center}
\vspace{5.0 cm}
\begin{center}
{\bf Abstract}
\end{center}
{The tools of Kerr/CFT correspondence are applied to the Kerr black hole embedded in an axial external magnetic field. Its extremal near horizon geometry remains a warped and twisted product of $AdS_2\times S^2$. The central charge of the Virasoro algebra, generating the asymptotic symmetries of the near horizon geometry, is found. It is used to reproduce, via the Cardy formula, the Bekenstein-Hawking entropy of the magnetised Kerr black hole as the statistical microscopic entropy of a dual CFT. The presence of the background magnetic field makes available also a second dual CFT picture, based on the $U(1)$ electromagnetic symmetry, instead of the only rotational one of the standard non-magnetised Kerr spacetime.\\
A Meissner-like effect, where at extremality the external magnetic field is expelled out of the black hole, allows us to infer the value of the mass for these magnetised extremal black holes.\\
The generalisation to the CFT dual for the magnetised extreme Kerr-Newman black hole is also presented.}
\end{titlepage}








\section{Introduction}

One of the biggest puzzles in gravity is the microscopic description of the black hole entropy. From macroscopic arguments we know, since the works of Bekenstein and Hawking in the 1970's, that the black hole entropy is proportional to a quarter of its event horizon area. Of course the main difficulties in this field are related to  the lack of a consistent theory of quantum gravity. Anyway in recent years some progress, inspired by the AdS/CFT conjecture \cite{maldacena}, but not relying on it, has been made to take into account the four-dimensional black hole microscopic degrees of freedom. In fact in \cite{strom08} a new duality between extremal Kerr black holes and two-dimensional Conformal field theory (CFT) was proposed, the Kerr/CFT correspondence, and then extended to more general black holes, such as the extremal (A)dS-Kerr-Newman one \cite{stro-duals}. It allows one to reproduce the Bekenstein-Hawking entropy of the black hole as the statistical entropy of a dual two-dimensional CFT, thanks to the Cardy formula. The extremality plays an important role because, at this special point of the parametric space, symmetries are enhanced. In fact extremal black holes can be considered, from thermodynamic point of view, as a ground state because their Hawking temperature is null and because there is no physical process that allows to reach an extremal state from a non-extremal one, which are considered excited states. Nevertheless several black holes are phenomenologically observed with angular momentum and mass ratio very close to extremality.  Some attempts to extend the correspondence outside extremality have been made, see \cite{cargese} or \cite{compere-review} for reviews, but usually at the price of introducing extra assumptions.  \\
The fundamental point in the Kerr/CFT correspondence is the emergence of the $U(1)\times SL(2,\mathbb{R})$ symmetry in the near horizon geometry. Recently the appearance of that symmetry in the near horizon region has been discovered also for extremal black holes embedded in an external magnetic field \cite{magn-rn}, specifically the magnetised Reissner-Nordstr\"om spacetime. Thus it was possible to borrow the tools of the Kerr/CFT correspondence to obtain the macroscopic black hole entropy from the statistical entropy of a CFT dual model. \\
In this letter, as announced in \cite{magn-rn}, we want to extend the results obtained for the magnetised Reissner-Nordstr\"om metric to the Kerr(-Newman) black hole embedded in an external magnetic field. The inclusion of external magnetic field is astrophysically significant because strong axial magnetic fields are actually measured at the center of galaxies \cite{nature}. They are caused by matter rapidly spinning in the accretion disk of the black hole. On the other hand, from  a theoretical point of view, it would be interesting to test how much it is possible to stretch the Kerr/CFT correspondence analytically deforming the black hole by extra parameters. Moreover the fact that the Kerr/CFT technique focuses mainly on the near horizon properties of the black  hole can circumvent the difficulties in the thermodynamical description  related to the asymptotically non-constant behaviour of these magnetised metrics.    \\
Finally the study of the near horizon limit can suggest a way to characterise  the conserved charges for black holes immersed in external magnetic field. In particular we are interested in the value of the mass, which still remains an open question \cite{booth}. \\
In the next section some of the basic properties of the magnetised Kerr (MK) black holes, useful for the rest of the paper, will be reviewed. In section \ref{NHEMK} the near horizon geometry of the extremal MK will be studied, then in section \ref{micro-entropy} the microscopic entropy of the black hole is calculated.  In section \ref{meissner}, thanks to a near horizon isometry between the magnetised and unmagnetised black hole, at extremality, we can deduce a value for the Kerr black hole embedded in an external magnetic field. The generalisation of all these results to the magnetised Kerr-Newman black hole is discussed in  appendix \ref{app1}.\\

\section{Magnetised Kerr Black Hole Review}
\label{Review}

The magnetised Kerr metric is a stationary and axisymmetric electrovacuum solution which describes a rotating black hole embedded in an external Melvin-like magnetic field. Asymptotically it does not have constant curvature, but the metric approaches, for large radial distances, the Melvin magnetic universe\footnote{Note that the electromagnetic field approaches the Melvin one only up to a duality transformation \cite{hiscock}, \cite{gibbons}.}. The metric and the electromagnetic potential can be generated, from the Kerr spacetime as a seed, by means of a Harrison transformation and the Ernst technique \cite{ernst-magnetic}, \cite{ernst-wild}. Its form is given by

\beq \label{MKerr}
         ds^2  =   -\frac{f}{|\L|^2}(\Delta_\varphi d\tilde{\varphi}- \omega d{\tilde{t}})^2 + \frac{|\L| ^2}{f} \left[ \r^2 d{\tilde{t}}^2 - e^{-2\g} \left( \frac{d\tilde{r}^2}{\Delta_r} + \frac{dx^2}{\Delta_x} \right) \right]    \quad ,
\eeq
\beq \label{MA}
          A = A_{\tilde{t}} d{\tilde{t}} +A_{\tilde{\varphi}} d\tilde{\varphi} \quad ,
\eeq
with
\bea
        &&     \Delta_r(\tilde{r})  :=  \tilde{r}^2 + a^2-2m\tilde{r}         \quad  , \qquad            \D_x(x)  :=  1-x^2     \qquad ,  \qquad  \Delta_\varphi :=1 + a^2 m^2 B^4 \  , \quad  \\
         &&       f(\tilde{r},x):= - \frac{h \ \D_x}{\Sigma}   \quad  \  \qquad   ,    \qquad     \Sigma(\tilde{r},x):= r^2+a^2 x^2 \qquad ,\qquad    \rho^2 (\tilde{r},x)  := \Delta_r \D_x      \   \quad  ,    \\
         &&  h(\tilde{r},x):= (\tilde{r}^2 + a^2)^2  - a^2 \D_r \D_x   \quad \  ,        \qquad       \omega(\tilde{r},x) := \frac{a(1-a^2 m^2 B^4) - \b \D_r }{\tilde{r}^2+a^2} + \frac{3}{4} a m^2 B^4            \quad ,       \\
         &&   \L(\tilde{r},x) := 1 + \frac{B^2}{4} \left[ (\tilde{r}^2+a^2)\D_x - 2 i a m x (3-x^2) + \frac{2 m a^2 \D_x^2}{\tilde{r} + i a x} \right]   \quad , \qquad    e^{-2\g}(\tilde{r},x) := h \ \Delta_x     \qquad    \\
         &&    \beta(\tilde{r},x) := \frac{a \Sigma}{h} + \frac{B^4}{16} \left\lbrace -8 m \tilde{r} a x^2 (3-x^2) - 6 m \tilde{r} a (1-x^2)^2 + \frac{2ma^3(1-x^2)^3}{h} \left[ (\tilde{r}^2 + a^2) \tilde{r} +2 m a^2  \nn \right] \right.   	 \\      
         && \left. \qquad \quad \  + \frac{4 m^2 a^3 x^2}{h}  \left[ (\tilde{r}^2+a^2)(3-x^2)^2 - 4 a^2 (1-x^2) \right] \right\} \\
         &&  A_{\tilde{\varphi}}(\tilde{r},x) := A_{\tilde{\varphi}_0} + \frac{B \D_\varphi}{8 \Sigma |\L |^2}  \left\{ a^6 B^2 x^2 (1-x^2)^2 +\tilde{r}^4 (1-x^2)[4+B^2 \tilde{r}^2(1-x^2)] \right.  \nn \\
         && \  \ \quad +   a^2 \tilde{r} \Big[ 4 B^2 m^2 \tilde{r} x^2 (3-x^2)^2+4m(1-x^2)^2 [2+B^2\tilde{r}^2 (1-x^2)] + \tilde{r} [4-4x^4+B^2\tilde{r}^2(2-3x^2+x^6)] \Big]   \nn     \\
           && \left. \quad \ \  + a^4 \Big[ 4 x^2 (1-x^2) + B^2 [4 m \tilde{r} (1-x^2)^3 + 4m^2(1+x^2)^2 + \tilde{r}^2(1-3x^4+2x^6) ]\Big] \right\} \\ 
          &&  A_{\tilde{t}} (\tilde{r},x):=  A_{\tilde{t}_0} - \om \left( \frac{A_{\tilde{\varphi}} - A_{\tilde{\varphi}_0} }{\D_\varphi}\right) + \nn \\
         && \quad \qquad \ - a m B^3 \left[2\tilde{r} + \frac{1}{h} \left( 4 a^2 m^2 \tilde{r} - \tilde{r}(\tilde{r}^2+a^2) \D_r \D_x - \frac{1}{4} (\tilde{r}^3 -3 a^2 \tilde{r} + 2 m a^2) \D_r \D_x^2  \right) \right] ,\qquad    \label{At}
\eea
where the $m, a, B$ constants parametrise the mass, the angular momentum of the black hole and the external magnetic field respectively. The constant factor $\D_\varphi$ is explicitly added to avoid conical singularities on the symmetry axis $(\tilde{r}=0, x=\pm 1)$, without changing the usual range of the azimuthal angular coordinate $\tilde{\varphi} \in [0,2\pi)$. The other angular coordinate is $x = \cos \theta   \in [-1,1]$.\\ 
When the external magnetic field is vanishing, i.e. $ B = 0 $,  the solution (\ref{MKerr})-(\ref{At}) exactly recovers the Kerr black hole, while for vanishing $m$ and $a$ it gives the Melvin magnetic universe. In order to ensure the regularity of  the electromagnetic field on the rotation axis, $A_{\tilde{\varphi}_0}$ have to be gauge fixed such that  $ A_{\tilde{\varphi}} = 0 $ at $ x = \pm 1 $:
\beq \label{gauge-A}
            \bar{A}_{\tilde{\varphi}_0} = - 2 a^2 m^2 B^3   \quad .
\eeq
Exactly, as for the standard Kerr black hole, the magnetised solution posses an inner $\tilde{r}_-$ and an outer (event) horizon $\tilde{r}_+$ located at
\beq
                    \tilde{r}_\pm = m \pm \sqrt{m^2-a^2} \ \quad .
\eeq 
In this letter we are mainly interested in the extremal limit of the  magnetised Kerr solution (\ref{MKerr})-(\ref{At}), (henceforth EMK) which is given, as in the case without the external magnetic field, in the limit $m \rightarrow a$, therefore the inner and outer horizon coincide at a double degenerate event horizon $\tilde{r}_e=a$. \\
The geometry of the magnetised black hole is modified, by the presence of the external magnetic field, with respect to the ordinary Kerr black hole. In fact the horizon of the MK is stretched along the direction of the magnetic flux. This can simply be understood from the ratio between the black hole area
\beq \label{area}
     \mathcal{A} = \int_0^{2\pi} d\tilde{\varphi} \int_{-1}^1 dx  \sqrt{g_{\tilde{\varphi}\tilde{\varphi}}g_{xx}} \Big|_{\tilde{r}=\tilde{r}_+} = 4 \pi \D_\varphi (\tilde{r}^2_+ +a^2)
\eeq
and the equatorial $C_{eq}$ circumference of the event horizon (evaluated at constant ${\tilde{t}}$ and $x=0$)
\beq
      C_{eq} =  \int_0^{2\pi} \sqrt{g_{\tilde{\varphi}\tilde{\varphi}}} \ d\tilde{\varphi} \Big|_{\tilde{r}=\tilde{r}_+} = \frac{4 \pi \D_\varphi m}{1+B^2 m^2} \quad .
\eeq
In fact, as the intensity of the external magnetic field gets stronger, the area grows while the equatorial circumference shrinks. Thus the event horizon shape becomes oblate.  \\
The extreme black hole area  is given by
\beq  \label{Aext}
          \mathcal{A}^{ext}=\lim_{m \rightarrow a} \mathcal{A} = 8 \pi a^2 \D_\varphi^{ext} \quad ,
\eeq
where $  \D_\varphi^{ext} := \lim_{m \rightarrow a} \D_\varphi = 1+ a^4 B^4$ .\\
The interaction of the external magnetic field with the rotating Kerr black hole gives rise to non-trivial effects. First of all, in analogy with the Faraday induction, even though the starting seed metric is electrically uncharged, after the Harrison transformation the black hole acquires an intrinsic electric charge. It can be quantified by  
\beq \label{Q}
        Q =\frac{1}{8\pi} \int_{\mathcal{S}_t} F^{\m\n} dS_{\m\n} = - \frac{1}{4\pi} \int_0^{2\pi} d\tilde{\varphi}  \int_{-1}^1 dx \ \sqrt{g_\mathcal{S}} \ n_\m \s_\n F^{\m\n}  =  2 a m B   \quad ,
\eeq
where $ dS_{\a\b} = - 2 n_{[\a}\s_{\b]} \sqrt{g_\mathcal{S}} \ d\tilde{\varphi} dx $, with $ \sqrt{g_\mathcal{S}} = \sqrt{g_{xx} g_{\tilde{\varphi} \tilde{\varphi}}} $, is the infinitesimal volume element of the two-dimensional integration surface $\mathcal{S}_t$ (for fixed time and radial coordinates) surrounding the black hole horizon. $n_\m$ and $\s_\n$ are two orthonormal vectors, time-like and space-like respectively, normal to the surface of integration $ \mathcal{S}_t $. On the other hand the magnetic charge $ P = \frac{1}{4\pi} \int_\mathcal{S} F $  remains null  because the Harrison transformation,  acting on the Kerr seed, does not generate magnetic monopoles.  Hence space-time is endowed only with an external magnetic field, whose amount of entering flux lines, inside the surface $\mathcal{S}$, is equal to the outgoing ones.   \\
A second non-trivial effect is the contribution of the external magnetic field on the angular momentum. Using the method of \cite{barnich}, for computing conserved charges, we can reproduce the result of \cite{gibbons2} and \cite{booth}, but only if we consider the electromagnetic potential to be properly regularised  by the gauge fixing (\ref{gauge-A}): 
\bea \label{J}
       J &=& \frac{1}{16\pi} \lim_{\mathcal{S}_t\rightarrow \infty} \int_ {\mathcal{S}_t} \left[ \nabla^\a \xi^\b_{(\varphi)} + 2 F^{\a\b} \xi^\m_{(\varphi)} A_\m \right] dS_{\a\b}   \\
       &=& am \left( 1 + 2 A_{\tilde{\varphi}_0} B + 3 a^2 m^2 B^4 \right)  \Big|_{ A_{\tilde{\varphi}_0} = \bar{A}_{\tilde{\varphi}_0}} = am \left(1 - a^2 m^2 B^4 \right)  \ \ , \quad 
\eea
where $\xi^\m_{(\varphi)}$ represents the rotational Killing vector $\p_{\tilde{\varphi}}$.  \\
In the rest of the paper it will be useful to know the  electrostatic Coulomb potential $\Phi_e$ of the magnetised solution, and its extremal limit $\Phi^{ext}_e$. They are defined on the horizon as
 \bea \label{Phi}
               \Phi_e := - \chi^\m A_\m\Big|_{\tilde{r}=\tilde{r}_+}   &=&  -A_{\tilde{t}_0} + \frac{mB}{a} \frac{ 4m + a^2 B^2 (m+4B^2 m^3 -4\sqrt{m^2-a^2}) }{1+a^2 m^2 B^4} \qquad , \\
            \label{Phi-ext}   \ \Phi_e^{ext} :=   \lim_{m \rightarrow a} \Phi_e  \ &=&    - A_{\tilde{t}_0} + 2 a B + \frac{a^3 B^3}{2+2a^4B^4}  \quad ,
 \eea
where the Killing vector $\chi:= \p_t + \Omega_J \p_\varphi$ is the tangent to the horizon's null generator,
and  (\ref{gauge-A}) have been used. 
The chemical potential $\Omega_J$, conjugate to the angular momentum $J$, represents the angular velocity of the black hole horizon, it is given by
\beq \label{Omega}
                \Omega_J = -\frac{g_{\tilde{t} \tilde{\varphi}}}{g_{\tilde{\varphi}\tilde{\varphi}}} \ \Bigg|_{\tilde{r}=\tilde{r}_+} = \frac{a-m^2 a^3 B^4 + \frac{3}{2} a \tilde{r}_+ m^3 B^4}{\D_\varphi (\tilde{r}^2_+ + a^2)} \quad .
\eeq
The angular velocity in the extremal limit, $ m \rightarrow a$, becomes
\beq \label{Om-ext}
       \Omega_J^{ext} = \frac{2+a^4 B^4}{4 a \D_\varphi^{ext}} \quad .
\eeq
\vspace{-0.4cm}

\section{Near horizon of the extremal magnetised Kerr spacetime}
\label{NHEMK}

Following \cite{Bardeen:1999px} and \cite{compere-review}, in order to analyse the region very near the extreme magnetised Kerr event horizon $\tilde{r}_e$, we define new dimensionless coordinates ($ t , r , \varphi $) in this way:
\beq
            \tilde{r}(r) := r_e+ \l r_0 r    \qquad  ,  \qquad      \tilde{t}(t) := \frac{r_0}{\l} t     \qquad  ,    \qquad     \tilde{\varphi}(\varphi,t):= \varphi + \Omega_J^{ext} \ \frac{r_0}{\l} t  \qquad  , \qquad
\eeq
where $r_0$ has been introduced to factor out the overall scale of the near-horizon geometry.
In the presence of the electromagnetic potential $A_\m$, before taking the near-horizon limit, it is necessary to make the gauge transformation
\beq
          A_{\tilde{t}} \rightarrow A_{\tilde{t}} + \Phi_e  \quad .
\eeq 
The near horizon, extreme, magnetised, Kerr geometry (NHEMK) is defined as the limit of the EMK for $\l \rightarrow 0$. 
Remarkably the NHEMK geometry can be cast in the general form of the near-horizon geometry of standard extremal rotating black holes. This form posses a $ SL( 2 ,\mathbb{R} ) \times U(1) $ isometry\cite{compere-review}, which is a warped and twisted product of $AdS_2 \times S^2$, given by\footnote{Obviously the function $\a(x)$ can be absorbed in a coordinate transformation of the polar angle, but it is usually left because it can turn out to be convenient, in some cases.}
\beq \label{near-metric}
           ds^2 = \G(x) \left[ -r^2 dt^2 + \frac{dr^2}{r^2} + \a^2(x) \frac{dx^2}{1-x^2} + \g^2(x) \ \big( d\varphi + \k r dt \big)^2 \right] \quad ,
\eeq 
where for this space-time $ r_0 = a \sqrt{2} $, as for the Kerr black hole, while  
\bea \label{fields1}
           \G(x) &=& a^2 \left[ (1 + a^2 B^2)^2 + (1 - a^2 B^2)^2 x^2 \right]     \qquad , \qquad \a(x)= 1    \qquad \qquad  \qquad \ \ \quad  , \qquad   \\
    \label{fields2}    \g(x)  &=& \frac{2 a^2 \D_\varphi^{ext} \sqrt{1-x^2}}{\G(x)}       \qquad    \qquad  \qquad \qquad   \ \quad , \qquad \k = \frac{1-a^4 B^4}{\D_\varphi^{ext}}    \qquad . \quad 
\eea
The near horizon electromagnetic one form is given by
\beq \label{near-A}
               A = \ell(x) (d\varphi+\k r dt) - \frac{e}{\k} d\varphi \qquad ,
\eeq
where
\beq
           \ell(x) = -\frac{B \D_\varphi^{ext} r_0^2}{1-a^4B^4} \left[ 1 - \frac{2(1+a^2B^2)^2}{(1+a^2B^2)^2+(1-a^2B^2)^2 x^2} \right] \qquad ,  \qquad e = 2 a^4 B^3 - \bar{A}_{\tilde{\varphi}_0} \k  \quad .
\eeq
This important aspect of the near horizon geometry is coherent with the results of
 \cite{kunduri}, where it is shown that, for the Einstein-Maxwell theory we are considering, any extremal black hole near horizon geometry has the form (\ref{near-metric}). This fact opens up the possibility of further generalisations of the correspondence by additional deformations of the MK spacetime. \\
The $SL(2,\mathbb{R}) \times U(1)$ isometry of the NHEMK is infinitesimally generated, once the normalisation is fixed, by the following Killing vectors
\bea \label{killing}
              \z_{-1} = \p_t  \qquad &,& \qquad \z_0 = t \p_t - r \p_r \qquad \\
              \z_1 = \left( \frac{1}{2r^2} +\frac{t^2}{2} \right) \p_t - t r \ \p_r - \frac{\k}{r} \p_\varphi \qquad  &,& \qquad L_0 = \p_\varphi \qquad .          
\eea
From their commutation relations
\bea
              [ \z_0 , \z_{\pm} ] = \pm \z_{\pm}    \qquad \qquad , \qquad \qquad [ \z_{-1} , \z_1] =  \z_0  \qquad   ,     
\eea 
we can see that $\{\z_{-1}, \z_0, \z_{1}\}$ form the $SL(2,\mathbb{R}) \sim SO(2,1)$ algebra, while $L_0$ constitutes the $U(1)$ algebra.  \\
The fact that the external magnetic field does not break the black hole near horizon symmetry is the fundamental point in the microscopic analysis of the entropy, as will be done in the next section \ref{micro-entropy}. \\
According to the Kerr/CFT correspondence the thermodynamic properties of the extremal black holes are encoded in the asymptotic symmetry group of the near horizon geometry. Of course to properly define this group of asymptotic symmetries, the specification of the boundary conditions, for the metric and the electromagnetic potential, are fundamental. We can use the standard ones proposed in \cite{stro-duals}, (for a review see \cite{compere-review},\cite{magn-rn}) for the theory we are considering.\\
These boundary conditions are preserved by the following asymptotic killing vectors
\bea 
        \z_\epsilon &=&  \epsilon(\varphi) \p_\varphi -r \epsilon'(\varphi)  \p_r \  +\  \textrm{subleading terms} \quad  , \quad  \label{zep}\\
        \xi_\epsilon &=& -\left[\ell(\theta)-\frac{e}{\k} \right]  \epsilon(\varphi ) \   + \ \textrm{subleading terms}  \quad . \quad    \label{xiep}
\eea  
These two vectors obey the Virasoro algebra without the central extension, as can be seen by their Fourier mode expansion. \\

\section{Microscopic entropy}
\label{micro-entropy}

The appearance of the the Virasoro asymptotic algebra constitutes the first step towards the realisation of the duality between bulk quantum gravity on the NHEMK region and a two-dimensional boundary CFT. The next step would be the extension of the the Virasoro algebra of the near horizon asymptotic symmetries with the central charge. To do so it is necessary to consider the Dirac bracket between the symmetry generators. For the general metric described by (\ref{near-metric}) and for the theory we are considering, it is possible to prove \cite{compere-review} that matter does not contribute directly to the value of the central charge, but only influence it through the functions $\G(x),\g(x),\a(x)$ and $\k$, in the following  way
\beq \label{cc}
         c_J = \frac{3\ \k}{G_N \hbar} \int_{-1}^1 \frac{dx}{\sqrt{1-x^2}} \ \G(x) \a(x) \g(x) \quad .
\eeq 
In particular substituting in (\ref{cc}) the EMK fields (\ref{fields1})-(\ref{fields2}) and integrating we obtain the actual value of the central charge for the extremal magnetised Kerr black hole\footnote{In units where $G_N=1, \hbar=1$.}
\beq \label{CentralCharge}
          c_J = 12 a^2  (1 - a^4 B^4)    \quad .
\eeq
Making use of the hypothesis that the near horizon geometry of the extremal Kerr black hole embedded in a Melvin-like external magnetic field can be described by the left sector of a two-dimensional CFT, we apply the Cardy formula
\beq \label{cardy}
             \mathcal{S}_{CFT} = \frac{\pi^2}{3} (c_L T_L + c_R T_R)
\eeq
to compute the microscopical entropy of the dual systems. This formula is able to take into account some universal aspects of the thermodynamic behaviour of the CFT without the need of a detailed description of the theory. We stress that in the Kerr/CFT correspondence the applicability of the Cardy formula is an assumption, because the exact details of the CFT are not known. Therefore we can not know if the dual CFT falls into the class of applicability of the Cardy formula. A sufficient condition for its validity consists in the fact that the temperature is large compared with the central charge, or at least the temperature is larger than the lightest excitation levels of the theory, so that a large number of degrees of freedom are excited. In this sense the presence of the external magnetic field  improves the plausibility in the application of the Cardy formula because, as $B$ get closer to $\pm 1/a$,  we have both the desired effects: a smaller central charge and a bigger temperature, as can be seen from (\ref{CentralCharge}) and (\ref{Tphi}) respectively.   \\
For the moment we are considering only the rotational excitations along the direction generated by the Killing vector $\p_\phi$, corresponding to the left sector, therefore the right modes will considered frozen, so their temperature $T_R$ will vanish. As left temperature we cannot associate the Hawking temperature  because it is null for extremal black holes, as also the surface gravity, because the inner and outer horizon coincide, as can be seen in (\ref{th}). But still quantum states, just outside the horizon, are not pure states when one defines the vacuum using the generator of the horizon,as in \cite{strom08}, \cite{stro-duals}. 
In the  context of quantum field theory on curved background, to take into account these rotational 
 degrees of freedom, which generalise the Hartle-Hawking vacuum (build for the Schwarzschild black hole),the Frolov-Thorne temperature is used. It depends on the metric and matter fields but not directly on the theory. Generally it is defined  as the limit to extremality, that is $\tilde{r}_+\rightarrow \tilde{r}_e$ (or in our case $m \rightarrow a$)
\beq \label{Tphi}
            T_\varphi :=   \lim_{\tilde{r}_+\rightarrow \tilde{r_e}}  \frac{T_H}{\Omega_J^{ext}-\Omega_J}  = \frac{\D_\varphi^{ext}}{2 \pi (1 - a^4 B^4)} = \frac{1}{2 \pi \k }   \qquad ,
\eeq
where the Hawking  temperature $T_H$ is given in terms of the surface gravity $k_s$ as follows:
\beq \label{th} 
            T_H  :=  \frac{\hbar \ k_s}{2 \pi} =  \frac{\hbar}{2 \pi} \sqrt{-\mezzo \nabla_\m \chi_\n \nabla^\m \chi^\n} = \frac{\hbar}{2 \pi}  \ \frac{\tilde{r}_+ - \tilde{r}_-}{2 \tilde{r}_+^2} \qquad .
\eeq
The surface gravity remains exactly the same as the Kerr metric and the Frolov-Thorne temperature is of the same form (but different value) as the standard non-magnetised black holes, $T_\varphi =(2 \pi \k)^{-1}$. \\
Then using as left temperature $T_\varphi$ (\ref{Tphi}) in the Cardy formula (\ref{cardy}) we obtain the microscopic entropy of the EMK black hole
\beq \label{entropy}
           \mathcal{S_{CFT}}=\frac{\pi^2}{3} c_L T_L = 2 \pi a^2 \D_\varphi^{ext} = \frac{1}{4} \mathcal{A}^{ext}
\eeq
Note that the entropy of the dual CFT system coincides with a quarter of the extreme black hole area (\ref{Aext}), as happens for the standard Bekenstein-Hawking  entropy. Although the electromagnetic field is highly non trivial, the entropy is matched precisely with the contribution of the gravitational central charge only. This fact is in full analogy with the (extreme) AdS-Kerr-Newman case where the contribution to the central charge of the gauge field is also null.\\
Note also that, when the external magnetic field vanishes, we recover the usual result for the entropy of the extremal Kerr black hole, and also, in the null external magnetic field limit, all the intermediate steps remain meaningful. \\
However when the Kerr black hole is coupled with an electromagnetic field, such as the one which generates the external magnetic field, an additional $U(1)$ gauge symmetry is present besides the rotational one, around the azimuthal axis. This opens to the possibility to describe the EMK black hole in an alternative dual CFT way. In fact we can uplift, as in Kaluza-Klein theories, the electromagnetic potential as a compact extra dimension, with period $2\pi R_\psi$. It defines a $S^1$ gauge fibre, which, in higher dimensions, can be considered an angular coordinate parametrised by $\psi$. As explained in \cite{compere-review} it is possible to define a chemical potential associated with the direction generated by $\p_\psi$. Since the extra-dimensional angular coordinate has a period of $ 2 \pi R_\psi $, the extremal Frolov-Thorne temperature is expressed in units of  $R_\psi$
\beq  \label{Tpsi}
             T_\psi = T_e R_ \psi = \frac{R_\psi}{2 \pi  e}    \quad .
\eeq 
The electric chemical potential $ T_e $ can be defined, in analogy with $T_\varphi$, as  
\beq \label{Te}
   T_e :=   \lim_{\tilde{r}_+\rightarrow \tilde{r}_e}  \frac{T_H}{\Phi_e^{ext}-\Phi_e} = - \frac{\D_\varphi^{ext}}{2 \pi \left[ \bar{A}_{\tilde{\varphi}_0} (1-a^4 B^4) -2 a^4 B^3 \D_\varphi^{ext} \right]} \qquad ,
\eeq
where the electrostatic potentials $\Phi_e$ and $\Phi_e^{ext}$  were defined in (\ref{Phi}). It is a non-trivial fact that $T_e$ in (\ref{Te}) for the Kerr solution in external magnetic field remains of the same form as the whole (AdS)-Kerr-Newman spacetime family \cite{compere-review} and of the magnetised Reissner-Nordstr\"om black hole \cite{magn-rn}
\beq \label{Tef}
         T_e = {1\over 2 \pi e} \quad .
\eeq 
The alternative CFT picture assumes that the rotational degree of freedom are immersed in a thermal bath with temperature $T_\psi$, thus we can identify the left sector of the dual CFT with a density matrix at temperature $T_\psi$. Again under the assumption that at extremality there are no right excitations modes, we can assign
\beq  \label{assi}
                              T_L = T_\psi  \qquad , \qquad T_R  = 0  \quad .
\eeq  
Of course the gauge fibre parametrised by $\psi$ becomes degenerate when the external magnetic field vanishes, i.e. for $B=0$, therefore in that case (which coincides with the standard Kerr/CFT correspondence), while the final result of the entropy remains valid, many intermediate quantities become singular.  
Thanks to the five-dimensional uplift the central charge $c_Q$ can be computed also in this case, as done for the Kerr-Newman black hole \cite{compere-review}, \cite{204}:
\beq \label{ccQ}
          c_Q =  \frac{3\ e}{R_\psi} \int_{-1}^1 \frac{dx}{\sqrt{1-x^2}} \ \G(x) \a(x) \g(x) =  \frac{12 e}{R_\psi} a^2 \D_\varphi^{ext} \quad .
\eeq
Substituting $c_Q$ as the left central charge into the Cardy formula (\ref{cardy}) and using the eqs. (\ref{Tpsi})-(\ref{ccQ}) we recover again the entropy for the EMK black hole $\mathcal{S_{CFT}}= 2 \pi a^2 \D_\varphi = \mathcal{A}^{ext}/4$, precisely as in (\ref{entropy}).   \\
Note that, also in this alternative picture, the sufficient condition for the applicability of the Cardy formula is fulfilled when $B\approx \pm 1/a$. In fact, for this parametric region, $e$ become small, assuring that $c_Q << T_e $.\\
In case we want to do a consistency check for the thermodynamics quantities above computed,  the first law of black hole thermodynamics is not the best option here. This because in the extremal case the Hawking temperature is null, so the contribution  of the entropy to the first law vanishes
\beq 
               T^{ext}_H \d S = 0 = \d M - \Omega_J^{ext} \d J  - \Phi_e^{ext} \d Q \quad .
\eeq
Moreover the mass value for these magnetised rotating black holes still remains an open issue \cite{booth}, also it is not clear how to find the values of the angular velocity $\Om_\infty$ and Coulomb potential $\Phi_\infty$ at spatial infinity \cite{gibbons2}, because naively they appear divergent for large values of the radial coordinate\footnote{This problem is related with a proper definition of a co-rotating observer at large value of the radial coordinate, for these magnetised rotating metrics \cite{gibbons}.}. Anyway, as explained in \cite{compere-review} and \cite{stro-duals}, considering  the extremal entropy as a function of the angular momentum and electric charge only\footnote{The external magnetic field, here, is not considered to vary independently, i.e. $B=B(J,Q)$.}, i.e. $S^{ext} = S^{ext}(J,Q)$, its variation can be quantified by
\beq\label{balance}
               \d S^{ext} = \frac{1}{T_\varphi} \d J + \frac{1}{T_e} \d Q   \quad , \quad
\eeq  
where the temperature conjugated to the angular momentum and electric charge are respectively defined as follows
\beq \label{conj-temp}
       \frac{1}{T_\varphi}  = \left( \frac{\p S^{ext}}{\p J}\right)_Q        \qquad  \quad , \qquad \quad    \frac{1}{T_e} = \left( \frac{\p S^{ext}}{\p Q}\right)_J  \quad . \quad
\eeq
It is easy to verify that, substituting  in (\ref{conj-temp}), as the extremal entropy,  the entropy we have obtained with the Cardy formula $\mathcal{S}_{CFT}$ (\ref{entropy}), expressed in therms of the conserved charges $(J,Q)$
$$ \mathcal{S}_{CFT}(J,Q) = \pi \sqrt{4 J^2 +Q^4} \ ,  $$ we  find perfect agreement with the Frolov-Thorne extremal temperatures of eqs. (\ref{Tphi}) and (\ref{Te})-(\ref{Tef}). Hence the balance equation \label{balance} is consistently fulfilled.\\
\\
All the analysis we have done here for the EMK solution hole can be straightforwardly extended also to a rotating extreme black hole endowed with intrinsic electric charge $q$, embedded in an external magnetic field, i. e. the extreme magnetised Kerr-Newman black hole (EMKN) \cite{ernst-wild}, which have been analysed in \cite{hiscock}, \cite{gibbons} and \cite{booth}.  
At the same time, this case also represents the generalisation to the microscopic entropy of the extreme magnetised Reissner-Nordstr\"om (EMRN) black hole \cite{magn-rn}. Even thought we leave the details in appendix \ref{app1}, we are giving the values of the central charge $\hat{c}_J$ and of the temperature $\hat{T}_\varphi$ which can be used to reproduce, from a microscopic point of view, the classical entropy area law in the general case
$$
         \hat{c}_J =  -\frac{3}{4} \left[ 32 q a^2 B^3 (a^2+q^2) + 4 B q^3 (4+B^2 q^2) +16 a \sqrt{a^2+q^2} ( a^4 B^4 + q^2 a^2 B^4 -1 + \frac{3}{2} B^2q^2 + \frac{3}{16} B^4q^4) \right] \nn 
$$
$$         
         \hat{T}_\varphi =  \frac{-4 (2a^2+q^2)\left( 1 + \frac{3}{2} B^2 q^2 + a^2 q^2 B^4 + 2 a q B^3 \sqrt{a^2+q^2}  + a^4 B^4 + \frac{B^4 q^4}{16}  \right)}{\pi\left[32qa^2B^3(a^2+q^2) + 4 B q^3 (4+B^2 q^2) + 16 a \sqrt{a^2+q^2} (a^4 B^4 +  q^2 a^2 B^4 -1+\frac{3}{2} B^2q^2 + \frac{3}{16} B^4q^4) \right]}
$$

\beq \label{SEMKN}
         \mathcal{S}_{EMKN} = \frac{\pi^2}{3} \hat{c}_J \hat{T}_\varphi = \pi (2 a^2 + q^2) \left[  1 + \frac{3}{2} B^2 q^2 + a^2 B^4 (a^2+q^2) + 2 a q B^3 \sqrt{a^2+q^2}  + \frac{B^4 q^4}{16}  \right] \ .
\eeq
Also in this case the area law for the entropy $\mathcal{S}_{EMKN} = \mathcal{A}_{EMKN}/4 $ is fulfilled, as it can be checked using the area of the magnetised KN black hole of eq (\ref{area-gen}) or  \cite{gibbons2}.\\   
It is easy to verify that the above expression of $\hat{c}_J$ and $\hat{T}_\varphi$ reduce to the ones of EMRN \cite{magn-rn} and EMK (\ref{Tphi}), (\ref{cc}), (\ref{entropy}) for $a=0$ and $q=0$, respectively. 

Unfortunately outside the extremal limit of magnetised black holes it is not clear if it is possible to extend the applicability of the Kerr/CFT correspondence, because some of the near horizon spacetime symmetries seem to be broken. In particular the wave equation for a probe scalar field in the background of the magnetised black hole is not separable, except in the weak magnetic field approximation, that is small $B$. It means that some tools available for the standard Kerr black hole can not be applied. \\

\section{Mass and Meissner effect for extremal magnetised black holes}
\label{meissner}

The remarkable coincidence we have found, in the extremal case, of the near-horizon symmetries between the magnetised Kerr black hole (also unveiled in \cite{magn-rn} for the magnetised Reissner-Nordstr\"om solution) and the unmagnetised Kerr-Newman metric, besides the results of \cite{booth} and \cite{bicak}, makes natural to deepen scrutinise a possible isometry relating these electrovacuum solutions, in the near horizon region. Physically it can be interpreted as a Meissner-like effect, originally known when a sample, in presence of an external magnetic field,  is cooled below its superconducting transition temperature. It that case the magnetic field in the interior of the sample vanish, while it increase outside. A similar effect  might arise in this magnetised black hole too, when their (Hawking) temperature reduces to the extremal one and the external magnetic field is expelled from the black hole, so the metric in the near horizon region become similar to the non-magnetised Kerr-Newman one.\\
From a mathematical point of view it means that the black hole parameters $a$ and $q$ can be rescaled to reabsorb the intensity of the external magnetic field, $B$. We will mostly focus on the  map between the near horizon extremal  magnetised  Kerr black hole, as  presented in eqs. (\ref{MKerr})-(\ref{At}), and the standard extreme Kerr-Newman solution which, in the near horizon limit is characterised by (\ref{near-metric}), (\ref{near-A}) and the fields:
\bea 
\G_0(x) &=& q^2 + a^2 (1+x^2)  \hspace{1.3cm} ,  \hspace{0.9cm}  \a_0(x) \ = \ 1   \quad   ,    \label{fields01}   \\
\g_0(x)  &=&    \frac{(2a^2+q^2)\sqrt{1-x^2}}{\G_0(x)} \qquad ,      \hspace{1cm}     \ell_0(x) \ = \ \frac{q}{\k_0} \left[\frac{q^2+ a^2 (1- x^2)}{\G_0(x)} \right]    \quad   ,        \label{fields02}   \\
\k_0   &=&    2 a \frac{\sqrt{a^2+q^2}}{2a^2+q^2}      \hspace{1.7cm}  ,    \hspace{1.5cm}     e_0 \ = \ \frac{q^3}{2 a^2+q^2}     \quad   .            \label{fields03}       
\eea
It easy to verify that the near horizon behaviour of the EMK black hole as described in eqs (\ref{fields1}), (\ref{fields2}) and (\ref{near-A}) can be reproduced from the near horizon extreme Kerr-Newman fields (\ref{fields01}), (\ref{fields02}), (\ref{fields03})  upon rescaling the values of the parameters in this way:
\bea
         q &\mapsto &    q_M =  2 a^2 B \quad  ,  \label{amqm} \\
         a    &\mapsto &  a_M = a (1-a^2B^2)  \quad . \nn 
\eea 
Hence, just to give an example, we present the relation between the extremal magnetised black hole near horizon field $\G(x)$ and the EKN one, $\G_0(x)$:
\beq \label{Gam}
              \G(x)=\G_0(x) \bigg|_{\substack{ a=a_M  \\ q = q_M}} =  q_M^2 + a_M^2 (1+x^2)  \quad .
\eeq
Similarly, thanks to the transformation (\ref{amqm}),  many quantities can be mapped from the EKN to EMK black hole, including the event horizon area, conserved charges such as the electric charge and the angular momentum
\bea
          Q_{EKN} = q &\mapsto &    Q_{EMK} =  Q_{EKN} \Big|_{\substack{ q = q_M}}  = \ q_M = 2 a^2 B \quad ,  \label{QEMK}\\
           J_{EKN} = a \sqrt{a^2+q^2}  &\mapsto&  J_{EMK} =  J_{EKN} \   \bigg|_{\substack{ a=a_M  \\ q = q_M}}  = \  a_M \sqrt{a_M^2+q_M^2}  = a^2(1-a^2B^2)^2 \quad . \label{JEMK}
\eea
Indeed (\ref{QEMK}) and (\ref{JEMK}) respectively coincide with the extremal values of the electric charge $Q$ and angular momentum $J$ computed in (\ref{Q}) and (\ref{J}).\\
Nevertheless there are some quantities, that cannot be mapped in this way, as the angular velocity of the horizon $\Om_J$ or $\Phi_e$, which are a key ingredient in the definition of the left  of the dual conformal models. From this point we understand that the generalisation of the Kerr/CFT correspondence for magnetised black hole is not just a trivial isometric mapping of the extreme Kerr-Newman results.\\  
Concerning the definition of the mass for magnetised black hole the situation is more subtle and goes behind the scope of this letter, indeed, due to the unconventional asymptotic, the definition of energy in these magnetised spacetimes is not clear, for a recent discussion see \cite{booth}. Anyway, for the extremal case, assuming that the above map between the conserved charges of EKN and EMK solutions holds also for the energy, we can infer a value for the mass of the extreme Kerr black hole immersed in a Melvin-like external magnetic field
\beq
           M_{EMK} = M_{EKN}  \Big|_{\substack{ a=a_M  \\ q = q_M}} =  \sqrt{a_M^2+q_M^2} =  a (1+a^2B^2)   \label{magn-kerr-mass}
\eeq
This value for the mass coincides with the one computed in \cite{booth}, with the isolated horizon approach \cite{ashtekar}, but not with the one of \cite{gibbons2}.
By construction, the extremality condition is fulfilled also for the conserved magnetised charges $M^2_{EMK} = Q^2_{EMK}+J^2_{EMK}$, but the the first law of black hole thermodynamics is not satisfied, unless modify it by adding an extra magnetic chemical potential, as done in  \cite{gibbons2}\footnote{Alternatively in \cite{booth} the first law is fulfilled by taking $\Omega$ and $\Phi$ as prescribed by the isolated horizon formalism \cite{ashtekar}.}. This is another manifestation of the non-triviality of the Kerr/CFT correspondence in presence of the the external magnetic field, which goes beyond the near horizon isometry between the EMK and EKN black holes. Otherwise the first law would also have been respected by the mapping (\ref{amqm}).  \\
On the other hand, considering fixed the intensity of the external magnetic field $B$, another operative procedure is proposed in \cite{gibbons2} to fulfil the first law. It consists in choosing ad  hoc values for the angular velocity and Coulomb potential at infinity, respectively defined $\Omega^{ext}_\infty$ and $\Phi^{ext}_\infty$, to satisfy the first law. According to (\ref{Om-ext}) and (\ref{Phi-ext}) and setting $A_{\tilde{t}_0}=0$, we find
\bea
                 \Omega^{ext}_\infty       &=&     \frac{1+a^2B^2-2a^4B^4}{a \D_\varphi^{ext}} \quad , \\
                  \Phi^{ext}_\infty             &=&     \quad\frac{a B}{2 \D_\varphi^{ext} } \left( 2 - a^2 B^2 + 4 a^4B^4 \right).
\eea
Of course this does not provide a consistency check, but maybe it could be of some help in the open issue related to the definition of a co-rotating observer and angular velocity at spatial infinity, for these magnetised stationary spacetimes.\\ 
Note that the parametric region with improved applicability of the Cardy formula, corresponding to $a\approx \pm B^{-1}$, as seen in the previous section, is mapped, according to (\ref{amqm}), in small values of $a_M$, which means also small values of the total angular momentum $J_{EMK}$. \\
This mapping, in the near horizon region, can be realised also between the extremal magnetised Reissner-Nordstr\"om (analysed in \cite{magn-rn}) and EKN solution. The general case about the extremal magnetised Kerr-Newman black hole mass  is presented in appendix \ref{app1}.\\

\section{Summary and Comments}

In this letter the extreme Kerr black hole embedded in an external magnetic field is studied in the framework of the Kerr/CFT correspondence.\\
The symmetries of the near horizon geometry have been found and none of them are  broken by the presence of the external magnetic field. Hence the near horizon metric remains a warped and twisted product of $AdS_2 \times S^2$, endowed with the $SL(2,\mathbb{R}) \times U(1)$ isometry. The technology of the Kerr/CFT correspondence has been exploited to map the gravitational system to a two-dimensional dual CFT model. The Frolov-Thorne horizon temperature and central extension of the symmetry algebra, characterising the asymptotic near horizon geometry, are found. Thus it was possible to use the Cardy formula to reproduce, from a microscopic point of view, the Bekenstein-Hawking entropy of the magnetised black hole. Note that all the issues about the magnetised spacetimes asymptotic where avoided, because the dual CFT picture is only based on the near horizon region of the black hole. This is coherent with the fact that the the black hole entropy is a characteristic property of the event horizon.\\
The presence of the external magnetic field improves the applicability of the Cardy formula, at least for certain ranges of the magnetic field intensity. Moreover the $U(1)$ symmetry related to the electromagnetic gauge field, as in the case of electric charge, makes available an alternative dual CFT model, which gives again the correct value for the entropy of the gravitational system. From an astrophysical point of view the generalisation to the external magnetic field has also some phenomenological interest.\\
Further generalisation of this holographic description to the presence of an intrinsic electric charge, i.e.  magnetised extreme Kerr-Newman black hole, are direct. But the non-extremal case, when the external magnetic field is present, seems more involved, because some symmetries are not manifest in that case. \\
Therefore the Kerr/CFT correspondence have shown to be solid enough to describe  extreme black holes deformed by the presence of an external electromagnetic fields, for future perspective it would be interesting test its robustness considering more general deformations.\\
The fact that near the extremal near horizon symmetries remain of the same form as the non-magnetised Kerr-Newman case is related to a sort of Meissner effect, recently discovered for this kind of black holes immersed in a strong magnetic field  \cite{bicak}. It consists of the expulsion of the magnetic field near the event horizon region, upon rescaling some quantities, as happens in superconductivity. It's likely that this property is inherited  by the dual CFT model, which can display some properties as seen in superconductor materials.\\
The study of this effect makes us infer the value of the mass for these extremal Kerr(-Newman) black holes embedded in an external magnetic field. \\

\section*{Acknowledgements}
\small I would like to thank Tim Taves and Cedric Troessaert for fruitful discussions. 
\small This work has been funded by the Fondecyt grant 11150569. The Centro de Estudios Cient\'{\i}ficos (CECs) is funded by the Chilean Government through the Centers of Excellence Base Financing Program of Conicyt. \\
\normalsize

 \appendix
 \section{Magnetised Kerr-Newman/CFT}
 \label{app1}

In this section we give some details about the derivation of the microscopic entropy, as stated in eq (\ref{SEMKN}), for the intrinsically electrically charged ($q \ne 0$) black hole embedded in an external magnetic field, that is the magnetised Kerr-Newman spacetime \cite{ernst-wild}, \cite{hiscock}, \cite{gibbons}, \cite{booth}.  Its extreme near horizon geometry and near horizon electromagnetic potential can be expressed\footnote{For the magnetised Kerr-Newman quantities we will use a hat to distinguish them from the non-intrinsically charged ones (q=0) of the rest of the paper.} respectively with (\ref{near-metric}) and (\ref{near-A}), where  $r_0=\sqrt{2a^2+q^2}$ and 
\bea \label{fields-q}
           \hat{\G}(x) &=& \frac{1}{16} \left\{ q^2 (4+q^2B^2)^2 + 16 B^2 q^4 x^2  + 16 a^6 B^4 (1+x^2) +  8 a^4B^2\left[ 4-4x^2+3B^2q^2(1+x^2) \right]   \nn \right. \\
                             &+&   a^2 \left[ -8B^2q^2(x^2-7) +16(1+x^2) + 9 B^4q^4 (1+x^2)\right]  \nn \\
                             &+& 8 a B q \sqrt{a^2+q^2} \left[ 4 -4x^2+B^2q^2(1+3x^2) + 4a^2B^2(1+x^2) \right]  \nn \\
           \hat{\a}(x)  &=& 1  \\
           \hat{\g}(x) &=&  \frac{(2 a^2+q^2) \hat{\D}_\varphi^{ext} \sqrt{1-x^2}}{\hat{\G}(x)}    \nn  \\
           \hat{\k}   &=&  -\frac{32 a^2 B^3 q (a^2+q^2) + 4 B q^3 (4+B^2q^2) + a \sqrt{a^2+q^2} \left[ 16 a^2 B^4(a^2+q^2) -16 +24 B^2q^2 + 3 B^4q^4\right] }{8 \hat{\D}^{ext}_\varphi (2a^2+q^2)} \nn \\
           \hat{\D}_\varphi &=& 1 + a^2 m^2 B^4 + \frac{3}{2} B^2 q^2 + \frac{1}{16} B^4 q^4 +2 a q m B^3 \qquad , \qquad \qquad \hat{\D}_\varphi^{ext} := \lim_{m \rightarrow \sqrt{a^2+q^2}} \hat{\D}_\varphi \nn   \\
            \hat{e} &=& \frac{1}{8 \hat{\D}_\varphi^{ext} (2a^2+q^2)}  \left\{ 2 B^2 q (2 \hat{\bar{A}}_{\tilde{\varphi}_0} B + 3 \D_\varphi^{ext} ) (8 a^4+8 a^2 q^2+q^4) + 8 q^3 (2 \hat{\bar{A}}_{\tilde{\varphi}_0} B + \D_\varphi^{ext} ) + \right. \nn \\
            & & \qquad \qquad \qquad \quad \ \ \ + \  a \sqrt{a^2+q^2} \left[  \hat{\bar{A}}_{\tilde{\varphi}_0} \left( 16 a^4 B^4+16 a^2 B^4 q^2+3 B^4 q^4+24 B^2 q^2-16\right) + \right. \nn \\
             & & \qquad \qquad \hspace{2cm} \qquad \qquad \left. \left. \ \ + \ 2B \D_\varphi^{ext} \Big( 16 a^4 B^2+16 a^2 B^2 q^2+3 q^2 \left(B^2 q^2+4\right) \Big)  \  \right] \ \right\}  \nn \\
           \hat{\bar{A}}_{\tilde{\varphi}_0} &=& - \frac{B}{8} \big( 12q^2 + B^2 q^4 + 24 a B q m + 16 a^2 B^2 m^2 \big) \label{gaugefixA}
\eea
In the limit of null intrinsic electric charge ($q\rightarrow0$)  these hatted quantities reduce to the ones of section \ref{NHEMK}, while for vanishing $a$ we exactly recover the results for the magnetised Reissner-Nordstr\"om case \cite{magn-rn}.\\
Thanks to the insight of section \ref{meissner}, we can make use of the mapping between the EKN and the  EMKN black holes defined  by 
\bea
         q &\mapsto &    \hat{q}_M = q + 2 a B \sqrt{a^2+q^2} - \frac{B^2}{4} q^3 \quad  ,  \label{hatamqm} \\
         a    &\mapsto &  \hat{a}_M = a - q  B \sqrt{a^2+q^2} - \frac{B^2}{4} a (4a^2+3q^2)   \nn 
\eea 
to express more briefly these near horizon fields (\ref{fields-q}) for the magnetised case, once regularised the azimuthal component of the electromagnetic potential through the gauge fix (\ref{gaugefixA});  as it was done in section \ref{meissner} for the magnetised Kerr case. 
As an example we present how to obtain the field $\hat{\G}(x)$, through to the map (\ref{hatamqm})
\beq
              \hat{\G}(x)=\G_0(x) \bigg|_{\substack{ a=\hat{a}_M  \\ q = \hat{q}_M}} =  \hat{q}_M^2 + \hat{a}_M^2 (1+x^2)  \quad .
\eeq
The transformation (\ref{hatamqm}) maps the intrinsic electromagnetic charge and the angular momentum of the unmagnetised extremal black hole to the magnetised one 
\bea
          Q_{EKN} = q &\mapsto &  Q_{EMKN} =  Q_{EKN} \Big|_{\substack{ q = \hat{q_M}}}  = \hat{q}_M = q + 2 a B \sqrt{a^2+q^2} - \frac{B^2}{4} q^3 \ \ ,  \label{QEMKN}\\
           J_{EKN} = a \sqrt{a^2+q^2}  &\mapsto&  J_{EMKN} =  J_{EKN}   \bigg|_{\substack{ a=\hat{a}_M  \\ q = \hat{q}_M}}  =   \hat{a}_M \sqrt{\hat{a}_M^2+\hat{q}_M^2}  =\label{JEMKN} \\   \nn
          && \qquad \quad =  a \sqrt{a^2+q^2} - q^3 B - \frac{3 B^2}{2}  a q^2 \sqrt{a^2+q^2} +\\
          & &\qquad \quad - \frac{B^3}{4} q \left( 8 a^4 + 8 a^2 q^2 + q^4  \right) - B^4 a \sqrt{a^2+q^2} \left( a^4 + a^2 q^2  + \frac{3}{16} q^4 \right) \ . \nn
\eea
These values coincide with the electric charge and angular momentum of the EMKN black hole directly computed, as analogously done in (\ref{Q}) and (\ref{J}) for $q=0$. For $q \ne 0$, using the magnetised Kerr-Newman metric \cite{ernst-wild}, regularised\footnote{A Mathematica notebook with this metric can be found at \url{https://sites.google.com/site/marcoastorino/papers/1508-01583}.} with the gauge fixed $ \hat{\bar{A}}_{\tilde{\varphi}_0}$, as in (\ref{gaugefixA}), and with the rescaled (by a factor  $\hat{\D}_\varphi$) azimuthal coordinate, we have
\beq \label{Qhat}
        \hat{Q} =\frac{1}{8\pi} \int_\mathcal{\hat{S}} \hat{F}^{\m\n} d\hat{S}_{\m\n} = - \frac{1}{4\pi} \int_0^{2\pi} d\tilde{\varphi}  \int_{-1}^1 dx \ \sqrt{g_\mathcal{\hat{S}}} \ \hat{n}_\m \hat{\s}_\n \hat{F}^{\m\n}  =   q + 2 a m B  - \frac{1}{4} q^3 B^2\quad ,
\eeq
\bea \label{Jhat}
       \hat{J} &=& \frac{1}{16\pi} \lim_{\mathcal{S}_t\rightarrow \infty} \int_ {\mathcal{\hat{S}}_t} \left[ \nabla^\a \xi^\b_{(\varphi)} + 2 \hat{F}^{\a\b} \xi^\m_{(\varphi)} \hat{A}_\m \right] d\hat{S}_{\a\b}   \\
       &=& ( q \hat{A}_{\tilde{\varphi}_0} + a m ) + \left( 2 a m \hat{A}_{\tilde{\varphi}_0} + \frac{q^3}{2} \right) B  + \left(\frac{9}{2} a m q^2 - \frac{q^3}{4} \hat{A}_{\tilde{\varphi}_0}  \right) B^2 + \left( 6 q m^2 a^2 - \frac{q^5}{2} \right) B^3 + \nn \\
       &+& \left(3 a^3 m^3 - \frac{11}{6} ma q^4 \right) B^4 + \left( -\frac{1}{2} a^2  q^3 m^2-\frac{q^7}{32} \right) B^5 \ \Bigg|_{ \hat{A}_{\tilde{\varphi}_0} =  \hat{\bar{A}}_{\tilde{\varphi}_0}} = \nn \\
        &=& am - q^3 B - \frac{3}{2} ma q^2 B^2 - B^3 \left(\frac{q^5}{4} + 2 q m^2a^2   \right) - B^4 \left( m^3 a^3 - \frac{3}{16} ma q^4 \right)  \ \ . \quad 
\eea
It's easy to check that (\ref{QEMKN}) and (\ref{JEMKN}) match with the extremal limit of (\ref{Qhat}), (\ref{Jhat}) and also with the standard literature results \cite{gibbons2}, \cite{booth}.   \\
Assuming that, as the above two conserved charges, also the mass of the black hole can be mapped according to (\ref{hatamqm}), we can deduce the value of the mass for these EMKN black holes:
\beq \label{mEMKN}
            M_{EMKN} = M_{EKN}  \bigg|_{\substack{ a=\hat{a}_M  \\ q = \hat{q}_M}} = \sqrt{\hat{a}_M^2+\hat{q}_M^2} =  \sqrt{a^2+q^2} + a q B + \frac{B^2}{4}   \sqrt{a^2+q^2}  \left(4 a^2+q^2 \right)
\eeq
We remark that this value (\ref{mEMKN}) precisely recover, in the extremal limit, the mass computed, by the isolated horizon approach, in \cite{booth} 
\bea
           M^2_{MKN} &=&  m^2 + 2 a q m B + \left( 2 m^2 a^2 -q^4 + \frac{3}{2} q^2 m^2 \right) B^2 +  \\
           &+& \left( 2 a q m^3 - \frac{3}{2} a q^3 m \right) B^3 + \left( a^2 m^4 - \frac{a^2}{2} m^2 q^2 +\frac{1}{16} m^2 q^4\right) B^4 \quad , \nn    \label{mMKN}
\eea
 but the mass (\ref{mEMKN}) does not agree with the one of \cite{gibbons2}. Unfortunately (\ref{mMKN}), unlike its extremal limit (\ref{mEMKN}) and all other conserved charges, even outside the extremality, cannot be written as, at most, a quartic polynomial in $B$. \\
The mapping (\ref{hatamqm}), coherently with section \ref{meissner}, tends to the one of  (\ref{amqm}) when $q\rightarrow0$.  But in this case, analysed in section \ref{meissner}, the analogy with the Meissner effect may be considered less appropriate because the map (\ref{amqm})   was between an EMK and an EKN spacetime. While for $q \ne 0$ the map (\ref{hatamqm}) relates an extremal Kerr-Newman black hole with its magnetised version.  \\ 
The Killing vectors of the near horizon geometry respect the isometric mapping (\ref{hatamqm}), thus they are of the same form of (\ref{killing}), but with the charged $\hat{\kappa}$ instead of $\k$. This can be considered a trivial consequence of the isometric mapping.\\
What seems less trivial, because non-extremal quantities are involved, is the fact that, following the same procedure of section \ref{micro-entropy},  for the charged near horizon geometry (\ref{near-metric}), (\ref{near-A}), (\ref{fields-q}), it is possible to find the Frolov-Thorne temperature in this more general setting. According to (\ref{Tphi}), we obtain    
\beq
            \hat{T}_\varphi :=   \lim_{\hat{\tilde{r}}_+\rightarrow \hat{\tilde{r}}_e}  \frac{\hat{T}_H}{\hat{\Omega}_J^{ext}-\hat{\Omega}_J}  = \frac{1}{2 \pi \hat{\k} }   \qquad ,
\eeq
where the inner and the event horizons are located at $\hat{\tilde{r}}_\pm = m \pm \sqrt{m^2-a^2-q^2}$, at extremality $ \hat{\tilde{r}}_+ $ goes to   $\hat{\tilde{r}}_e= \sqrt{a^2+q^2}$ and where
\beq
              \hat{T}_H   = \frac{1}{2 \pi} \frac{\hat{\tilde{r}}_+-\hat{\tilde{r}}_-}{2 (\hat{\tilde{r}}_+^2 + a^2)}  \quad ,
\eeq
\beq               \hat{\Omega}_J = \frac{ a - 2 B q \hat{\tilde{r}}_+ - \frac{3}{2} a B^2 q^2 - \frac{B^3}{2} q \Big[ 4a^2m-4m^2\hat{\tilde{r}}_+ q^2 (2m+\hat{\tilde{r}}_+)  \Big]  - \frac{B^4}{8} am \Big[ 8a^2m-12m^2\hat{\tilde{r}}_+ +3q(2m+\hat{\tilde{r}}_+) \Big] }{ \hat{\D}_\varphi (\hat{\tilde{r}}_+^2 + a^2)} \nn 
\eeq
Finally the central charge $\hat{c}_J$ can be computed from (\ref{cc}), using the near horizon function of the magnetised KN black hole (\ref{fields-q}); the result is given at the end of section (\ref{micro-entropy}). Otherwise, thanks to the mapping (\ref{hatamqm}), it can be shortly written as
\beq
                    \hat{c}_J = 6 \hat{\k}   \big( 2 \hat{a}_M^2 + \hat{q}_M^2 \big) \quad .
\eeq
Thus the entropy of the magnetised extreme Kerr-Newman black hole can be computed from (\ref{cardy}) 
\beq \label{entropy-hat}
                                 \hat{\mathcal{S}}_{CFT} = \pi  \big( 2 \hat{a}_M^2 +\hat{q}_M^2 \big) = \frac{1}{4} \hat{\mathcal{A}}^{ext} \quad ,
\eeq 
 as explicitly presented  in (\ref{SEMKN}). The area of the MKN and EMKN black holes is respectively given by
 \beq  \label{area-gen}
        \hat{ \mathcal{A}} = \int_0^{2\pi} d\tilde{\varphi} \int_{-1}^1 dx  \sqrt{\hat{g}_{\tilde{\varphi}\tilde{\varphi}} \hat{g}_{xx}} \ \Big|_{\tilde{r}=\hat{\tilde{r}}_+} = 4 \pi \hat{\D}_\varphi (\hat{\tilde{r}}^2_+ +a^2)  \quad , \qquad  \hat{ \mathcal{A}}^{ext}:=\lim_{m\rightarrow\sqrt{a^2+q^2}}  \hat{\mathcal{A}} \quad .
 \eeq
It can be shown that also the entropy (\ref{entropy-hat}) fulfil the balance equation (\ref{balance}), as in the case $q=0$.   \\
For this intrinsically charged case it is also possible to reproduce the entropy through the alternative CFT picture based on the $U(1)$ electromagnetic gauge symmetry, as described in section \ref{micro-entropy} for $q=0$, we leave the details for the reader. \\

\end{document}